\documentclass[useAMS,usenatbib]{mn2e}
\usepackage{amsmath,graphicx,txfonts}
\newcommand{\mr}{\mathrm}

\newcommand{\be}{\begin{equation}}
\newcommand{\ee}{\end{equation}}

\newcommand{\beq}{\begin{eqnarray}}
\newcommand{\eeq}{\end{eqnarray}}
\newcommand{\ensav}[1]{\left\langle #1 \right\rangle}
\def\lsim{\mathrel{\rlap{\lower4pt\hbox{\hskip1pt$\sim$}}
    \raise1pt\hbox{$<$}}}                
\def\gsim{\mathrel{\rlap{\lower4pt\hbox{\hskip1pt$\sim$}}
    \raise1pt\hbox{$>$}}}                
\def\aap{A\&A}
\def\apj{ApJ}
\def\apjl{ApJL}
\def\apjs{ApJS}
\def\mnras{MNRAS}

\def\physrep{Phys.~Rep.}   
\def\prd{Phys.~Rev.~D}        
\def\prl{Phys.~Rev.~Lett.}        
\def\pasj{PASJ}
\def\jcap{JCAP}

\begin{document}

\title{Tidal alignments as a contaminant of the galaxy bispectrum}
\author[E. Krause \& C. M. Hirata]{Elisabeth Krause$^1$ and Christopher M. Hirata$^2$
\\$^1$Caltech M/C 249-17, Pasadena, CA 91125, USA
\\$^2$Caltech M/C 350-17, Pasadena, CA 91125, USA
}
\date{\today}
\pagerange{\pageref{firstpage}--\pageref{lastpage}} \pubyear{2010}
\maketitle
\label{firstpage}
\begin{abstract}
If the orientations of galaxies are correlated with large-scale structure, then anisotropic selection effects such as preferential selection of face-on 
disc galaxies can contaminate large scale structure observables.  Here we consider the effect on the galaxy bispectrum, which has attracted interest 
as a way to break the degeneracy between galaxy bias and the amplitude of matter fluctuations $\sigma_8$.  We consider two models of 
intrinsic galaxy alignments: one where the probability distribution for the galaxy's orientation contains a term linear in the local tidal field, 
appropriate for elliptical galaxies; and one with a term quadratic in the local tidal field, which may be applicable to disc galaxies.  We compute the 
correction to the redshift-space bispectrum in the quasilinear regime, and then focus on its effects on parameter constraints from the 
transverse bispectrum, i.e. using 
triangles $(\bmath k_1,\bmath k_2,\bmath k_3)$ in the plane of the sky.
We show that in 
the linear alignment model, intrinsic alignments result in an error in the galaxy bias parameters, but do not affect the inferred value of $\sigma_8$.  
In contrast, the quadratic alignment model results in a systematic error in both the bias parameters and $\sigma_8$.  However, the quadratic alignment 
effect has a unique configuration dependence that should enable it to be removed in upcoming surveys.
\end{abstract}


\section{Introduction}

While the evolution of dark matter perturbations in the current $\Lambda$CDM model is well understood theoretically, the relation between the galaxy 
distribution and the large scale (dark) matter distribution is complicated by the detailed physics of galaxy formation and different models may lead to 
different clustering properties of galaxies.  In particular, while local theories of galaxy formation predict that the galaxy density fluctuations 
trace the matter fluctuations on large scales, they also predict that the two are related by the {\em bias parameter} $b$, which is in general not 
known {\em a priori} \citep{K84}.  The unknown bias parameter represents a key problem for attempts to measure the growth of cosmological perturbations 
using galaxies.

In combination with the galaxy power spectrum, third order galaxy clustering measures such as the bispectrum or (equivalently) the 3-point correlation 
function can be used to measure non-linear galaxy bias and break the degeneracy 
between the normalization of the matter power spectrum, $\sigma_8$, and the linear galaxy bias. This enables one remove the effects of galaxy biasing 
and measure the cosmological growth of structure from the galaxy distribution
\citep{F94, VHM98, SCF99, V00}, and thus constrain dark energy \citep[e.g.][]{DTJ06}. Recently 
third 
order galaxy clustering has been analyzed by several authors using the bispectrum \citep{SFF01,FFS01,VHP02,KNS07} and the three point correlation 
function \citep{JB04, KSN04,NSS06}. Using mock catalogs from numerical simulations, \citet{SCPS06} show that a combined analysis of the galaxy power 
spectrum and bispectrum including their cross-correlation contains significant information on galaxy bias and fundamental cosmological parameters and 
helps break parameter degeneracies of other cosmological probes.

The most important systematic errors in interpreting the observed galaxy clustering arise in the non-linear regime,where the behavior of galaxy biasing 
and models of the (redshift space) galaxy power spectrum and bispectrum are difficult to model \citep[see][for the complications of a current model of 
the redshift space Bispectrum]{S+08}. Recently \citet{Hirata09} showed that the alignment of galaxies by large scale tidal fields can cause a 
systematic 
error in the determination of the linear redshift space distortion parameter $\beta$ \citep{K87}: the 
alignment of galaxies with the tidal field (along the stretching axis of the field for large elliptical galaxies) in combination with a viewing 
direction dependent galaxy selection effect, e.g. preferential selection of galaxies which are observed along their long axis, will lead to a selection 
probability for galaxies which is modulated by the tidal field along the line of sight.  This results in an anisotropy in redshift-space clustering 
with the same scale and angular dependence as the linear redshift-space effect. 
In this paper we will explore the implications of such a tidal alignment contamination 
for the observed galaxy bispectrum and how it affects the measurement of galaxy bias parameters.

Throughout this paper we assume a standard $\Lambda$CDM cosmology with the best-fit WMAP 7 \citep{WMAP7} parameters, and assume Gaussian initial 
density perturbations.

We begin in Section~\ref{sec:theory} with a derivation of the standard redshift space galaxy bispectrum and discuss toy models for physical processes 
that cause alignments of galaxy orientations with large scale structure. In Section \ref{sec:TA} we explain how tidal alignments of galaxies in 
combination with an orientation dependent galaxy selection modify the observed galaxy distribution and calculate the corresponding corrections to the 
galaxy bispectrum. Using a Fisher matrix technique we then estimate the systematic error induced by tidal alignments to measurements of galaxy bias 
parameters from angular clustering in Section~\ref{sec:fisher}. We conclude and discuss mitigation strategies in Section~\ref{sec:theend}.

\section{Theoretical Background}
\label{sec:theory}

In this section we derive the redshift space galaxy bispectrum to second order in perturbation theory \citep[for a review, see e.g.][]{Bernardeau02}, 
and discuss toy models for the alignment of galaxies with the large scale tidal field.

\subsection{Galaxy bispectrum}

The matter bispectrum $B$ is defined as
\be
\ensav{\tilde{\delta}(\bmath k_1)\tilde\delta(\bmath k_2)\tilde\delta(\bmath k_3)}\equiv (2\pi)^3 \delta_{\mr D}\left(\bmath{k}_{123}\right)B(\bmath k_1,\bmath k_2,\bmath k_3)\;,
\ee
where $\tilde\delta(\bmath k)$ is the matter density contrast in Fourier space, $\delta_{\mr D}$ the Dirac delta function, and $\bmath k_{123} \equiv 
\bmath k_1+\bmath k_2+\bmath k_3$. The bispectrum vanishes for a Gaussian random field.

To second order perturbation theory the density contrast is given by 
\be
\tilde\delta(\bmath k) = \tilde\delta^{(1)}(\bmath k)+\int\frac{\mr d^3 \bmath k_1}{(2\pi)^3}F_2\left(\bmath k_1,\bmath k - \bmath 
k_1\right)\tilde\delta^{(1)}(\bmath k_1)\tilde\delta^{(1)}(\bmath k - \bmath k_1),
\label{eq:D2}
\ee
with $\tilde\delta^{(1)}({\bmath k})$ the linear density contrast, and the second order density kernel
\be
F_2(\bmath k_1, \bmath k_2) = \frac{5}{7}+\frac{\bmath k_1\cdot \bmath k_2}{2k_1 
k_2}\left(\frac{k_1}{k_2}+\frac{k_2}{k_1}\right)+\frac{2}{7}\left(\frac{\bmath k_1\cdot \bmath k_2}{k_1 k_2}\right)^2.
\ee
Hence the matter bispectrum induced by non-linear gravitational evolution at tree-level is given by
\be
B(\bmath k_1,\bmath k_2,\bmath k_3) = 2 F_2(\bmath k_1, \bmath k_2)P(k_1)P(k_2) + \mr{2~perm.},
\ee
where $P(k)$ is the linear matter power spectrum, $\bmath k_3 = -\bmath k_1-\bmath k_2$ and ``2 perm.'' indicates that the 2 permutations $(\bmath 
k_2,\bmath k_3)$ and $(\bmath k_1,\bmath k_3)$ are also included in the summation.

Using the local bias approximation \citep[e.g.][]{FG93}, the galaxy density contrast $\delta_{\mr g}$ can be expressed as a non-linear function of the 
matter density contrast
\be
\delta_{\mr g}(\bmath x) = b_1\delta(\bmath x)+\frac{1}{2}b_2\delta(\bmath x)^2+\cdots\;,
\label{eq:biasing}
\ee
where the expansion coefficients are the linear ($b_1$) and non-linear galaxy bias factors.  In reality, galaxy biasing may be more complicated, 
especially on small scales, due to 1-halo terms \citep{Seljak00} and nonlocal dependences such as the strength of the local tidal field 
\citep{McDonald, McDonald2}.
However, in simulations the local bias model is found to be a fair 
description of non-linear halo clustering on large scales with an accuracy of a few percent \citep[e.g.][]{Marin08,GJ09, MG09}, which is sufficient at 
the level of this analysis.

Then the galaxy bispectrum $B_{\mr g}$ is related to the matter bispectrum via
\be
B_{\mr g}(\bmath k_1,\bmath k_2,\bmath k_3)\simeq b_1^3 B(\bmath k_1,\bmath k_2,\bmath k_3)+b_1^2 b_2 \left(P(k_1)P(k_2)+\mr{2~perm.}\right),
\label{eq:B_g}
\ee
and similarly for the galaxy power spectrum,
\be
P_{\mr g}(k) = b_1^2 P(k).
\label{eq:P_g}
\ee

To arrive at an expression for the redshift space galaxy bispectrum we have to transform radial coordinates to redshifts space. In the plane-parallel 
approximation, the mapping from real space position $\bmath x$ to coordinate $\bmath x^{\mr s}$ in redshift space is given by
\be
\bmath x^{\mr s} = \bmath x + \frac{\hat{\bmath{n}}\cdot \bmath u (\bmath x)}{H a} \hat{\bmath{n}},
\label{eq:s}
\ee
where $\bmath u (\bmath x)$ is the peculiar velocity field, and 
$\hat{\bmath{n}} $ is the direction of the line of sight. The velocity field is curl-free, $\nabla\times\bmath u({\bmath x})=0$, at all orders in 
perturbation theory.  Its divergence is given to linear order in perturbation theory by
\be
{\rm i}{\bmath k}\cdot\tilde {\bmath u}^{(1)}({\bmath k}) = aHf\tilde \delta^{(1)}({\bmath k}),
\label{eq:f}
\ee
where $f =\mr d \ln(G)/\mr d \ln(a)$ is the logarithmic growth rate of linear perturbations (equal to roughly $\Omega_{\rm m}^{0.6}$ in general
relativity).  Higher-order contributions to $\nabla\cdot{\bmath u}$ \citep{Bernardeau02} are analogous to Eq.~(\ref{eq:D2}), e.g.
\be
{\rm i}{\bmath k}\cdot\tilde {\bmath u}^{(2)}({\bmath k}) = aHf \int \frac{{\rm d}^3{\bmath k_1}}{(2\pi)^3}
  G_2({\bmath k}_1,{\bmath k}-{\bmath k}_1)
  \tilde \delta^{(1)}({\bmath k}_1)\tilde \delta^{(1)}({\bmath k}-{\bmath k}_1),
\ee
with the kernel
\be
G_2(\bmath k_1, \bmath k_2) = \frac{3}{7}+\frac{\bmath k_1\cdot \bmath k_2}{2k_1 
k_2}\left(\frac{k_1}{k_2}+\frac{k_2}{k_1}\right)+\frac{4}{7}\left(\frac{\bmath k_1\cdot \bmath k_2}{k_1 k_2}\right)^2.
\label{eq:G2}
\ee

Taking into account the Jacobian of this mapping of $\bmath x\rightarrow\bmath x^{\rm s}$ (Eq.~\ref{eq:s}), and approximating the peculiar velocity field by 
the second order bulk  velocity field, the galaxy density is redshift space is \citep{HMV98,SCF99}
\beq
\tilde\delta_{\mr g}^{\mr s}(\bmath k^{\rm s}) 
\!\!\!\! &=& \!\!\!\! (b_1+f \mu^2)\tilde\delta^{(1)}(\bmath k^{\rm s})
\nonumber \\ && \!\!\!\!
+ \int\frac{\mr d^3 \bmath k_1^{\rm s}}{(2\pi)^3}Z_2\left(\bmath k_1^{\rm s},\bmath k^{\rm s} - \bmath 
k_1^{\rm s}\right)\tilde\delta^{(1)}(\bmath k_1^{\rm s})\tilde\delta^{(1)}(\bmath k^{\rm s} - \bmath k_1^{\rm s}),
\label{eq:dgs}
\eeq
where $\bmath k^{\rm s}$ denotes a Fourier mode in red shift space, and $\mu \equiv \hat{\bmath k}\cdot \hat{\bmath{n}}$ is the cosine of the angle between the wave vector and the line of sight (we may analogously 
define $\mu_1$, $\mu_{12}$, etc.).  The mode-coupling function $Z_2$ is
\beq
\nonumber Z_2(\bmath k_1, \bmath k_2) \!\!\! &=& \!\!\! b_1F_2(\bmath k_1, \bmath k_2) + f\mu_{12}^2 G_2(\bmath k_1, \bmath k_2)\\
&&\!\!\! +\frac{f\mu_{12}k_{12}}{2}\left[\frac{\mu_1}{k_1}(b_1+f \mu_2^2)+\frac{\mu_2}{k_2}(b_1+f \mu_1^2)\right]+\frac{b_2}{2},
\eeq
Hence we can write the redshift space galaxy bispectrum as
\beq
B_{\mr g}^{\mr s}(\bmath k_1^{\rm s},\bmath k_2^{\rm s},\bmath k_3^{\rm s}) \!\! &=& \!\! 2(b_1+f\mu_1^2)(b_1+f\mu_2^2) P(k_1^{\rm s})P(k_2^{\rm s})Z_2(\bmath k_1^{\rm s},\bmath k_2^{\rm s})
\nonumber \\ && \!\! + 2\,{\rm perm.}
\eeq
Note that this expression does not include the {\em Finger of God} effect due to the virialized motion of galaxies within a cluster \citep{Jackson72}, 
which is important when one of the $k_i$ has a large line-of-sight component.  While this effect 
is important even on weakly non-linear scales, it is usually handled by phenomenological models \citep[e.g.][]{HC98,VHM98,SCF99,P01}, a compression of 
radial 
coordinates for galaxies living in the same cluster \citep[e.g.][]{Tegmark04}, or by reconstructing the redshift-space halo density field \citep{Reid09}.

\subsection{Toy models of tidal alignments}

\subsubsection{Halo shape distortions: linear alignment}

In the linear alignment model \citep{CKB} the shape and orientation of a galaxy are assumed to be determined by the shape of the halo it resides in. It is 
thought that the gravitational collapse of an initially spherical overdensity in a constant gravitational field leads to triaxial haloes, such that the 
halo will be prolate if the overdensity is stretched by the large scale tidal field and oblate if it is compressed. This mechanism is believed to lead 
to a net correlation of halo orientations even though overdensities typically are not spherical, and such an alignment has been confirmed by 
simulations \citep[e.g.][]{FLW09}.

The relation between halo shape and galaxy shape is complicated by galaxy formation and differs between galaxy types \citep[e.g.][]{FLM07}, but at 
least for luminous red galaxies (LRGs) there is observational evidence for an alignment of the LRG with the major axis of its host \citep{B82,FLM07, 
NSD10}.  There are also correlations with large-scale structure \citep{B82}; with the Sloan Digital Sky Survey (SDSS) it has even been possible to 
measure the 
scale dependence of these correlations and show the consistency of their spectral index with the predictions of the linear tidal alignment model and 
the $\Lambda$CDM power spectrum \citep{H07}.

\subsubsection{Tidal torques: quadratic alignment}

The orientation of a disc galaxy is determined by the direction of its angular momentum, which builds up due to tidal torquing during early stages of 
galaxy formation if the proto-galaxy's inertia tensor is anisotropic and misaligned with the local shear field \citep{Hoyle,Sciama,P69,D70,W84,CNPT01}. 
See \citet{BMS09} for a 
review of tidal torquing and the build up of angular momentum correlations.

Following \citet{LP00}, we parameterize the correlation between moment of inertia and the shear field  by
\be
\ensav{L_i L_j} = \ensav{L^2}\left(\frac{1+\alpha}{3}\delta_{ij}-\alpha \hat T_{ih} \hat T_{hj}\right),
\label{eq:LiLj}
\ee
which is also the most general quadratic form possible. Here
$\hat T_{ij}$ is the unit normalised traceless tidal field tensor ($\hat T_{ij}\hat T_{ij}=1$) and $\alpha$ is a dimensionless coupling parameter, e.g.
$\alpha = \frac35$ at leading order 
in perturbation theory if shear and inertia tensor are mutually uncorrelated.  It is also possible for $\alpha$ to be much smaller, e.g. if the 
angular momentum vector of the disk is only partially aligned with that of the host halo \citep[e.g.][]{vandenBosch}.

Note that in non-linear theory spin-induced alignments also have a linear contribution at large scales because the large-scale tidal field induces 
correlations of the small-scale tidal field and inertia tensor that lead to a nonzero contribution to $\langle L_iL_j \rangle$
\citep{HZ08}, although this linear effect has not been observed for late-type galaxies despite several searches \citep{LP07, H07, MWiggleZ}.

\section{Tidal alignment contamination}
\label{sec:TA}

As discussed in the previous section, the orientation of galaxies likely is not random but correlated with large scale structure, and in combination with observational galaxy selection criteria which depend on the galaxy orientation relative to the line of sight, this may modify the observable galaxy distribution. Following \citet{Hirata09}, we will now introduce the basic notation needed to discuss galaxy orientation and viewing direction dependent selection effects.

Let the galaxy orientation be described by the Euler angles ($\theta,\phi,\psi$) through a rotation matrix $\mathbfss Q (\theta,\phi, \psi)$. This 
matrix transforms ``lab'' frame coordinates to a coordinate system aligned with the galaxy.
Due to tidal alignments the probability distribution $p(\mathbfss Q|\bmath x)$ for the orientation of a galaxy at position $\bmath x$ may be 
anisotropic and a function of the local environment of $\bmath x$.
The observational galaxy selection probability depends on the direction of the line of sight, $\hat{\bmath n}$, and the galaxy orientation, 
specifically on the direction of the line of sight in the galaxy frame $\mathbfss Q \hat{\bmath n}$.  We define
\be
P\propto 1 + \Upsilon\left(\mathbfss Q \hat{\bmath n},\bmath x\right),
\label{eq:Upsilon}
\ee 
where the anisotropic part $\Upsilon$ is zero when averaged over all possible galaxy orientations or viewing directions.

The observable galaxy distribution $N (\mr{selected})$ hence is modified compared to the true galaxy distribution $N(\mr{true})$ by
\beq
\nonumber \frac{N (\mr{selected})}{N(\mr{true})} (\hat{\bmath n}|\bmath x) \!\!&\propto&\!\! \int_{\mr{SO}(3)} p(\mathbfss Q|\bmath 
x)\left[1+\Upsilon\left(\mathbfss Q \hat{\bmath n},\bmath x\right)\right]\; \mr d ^3\mathbfss Q \\
&=&\!\! 1+ \int_{\mr{SO}(3)} p(\mathbfss Q|\bmath x)\Upsilon\left(\mathbfss Q \hat{\bmath n},\bmath x\right)\; \mr d ^3\mathbfss Q
\nonumber\\
&\equiv &\!\!
1+\epsilon(\hat{\bmath n}|\bmath x),
\label{eq:defepsilon}
\eeq
which is the average of Eq.~(\ref{eq:Upsilon}) over the distribution of galaxy orientations, and where we have defined the orientation dependent 
selection 
function $\epsilon(\hat{\bmath n}|\bmath x)$ in the last step. 
As the average of $\Upsilon$ over all galaxy orientations vanishes, Eq.~(\ref{eq:defepsilon}) implies that $\epsilon$ vanishes
if either the galaxy orientations are isotropically distributed or if the probability for selecting a galaxy is independent of $\mathbfss Q \hat{\bmath 
n}$, i.e. if $\Upsilon = 0$.

The observed galaxy density is modified by the orientation dependent selection function such that
\be
1 + \delta_{\mr g}^{\mr{obs}}(\bmath x^{\mr s})= \left(\left[1 +\delta_{\mr g}(\bmath x)\right]\left[1+\epsilon(\hat{\bmath n}|\bmath x)\right]\right)^{\mr s},
\ee
where the term in round brackets is the orientation modulated real space density of selected galaxies, and where the superscript $\mr s$ denotes the transform to redshift space. 
Expanding to second order in the matter density field, this implies:
\beq
\nonumber \tilde\delta_{\mr g}^{\mr{obs}}(\bmath k^{\rm s}) \!\!&=&\!\! \tilde\delta_{\mr g}^{\mr s(1)}(\bmath k^{\rm s})+ \tilde\epsilon^{\mr s (1)}(\hat{\bmath n}|\bmath 
k^{\rm s})+\tilde\delta_{\mr g}^{\mr s(2)}(\bmath k^{\rm s})+ \tilde\epsilon^{\mr s (2)}(\hat{\bmath n}|\bmath k^{\rm s})\\
&&\!\!+ \int\frac{\mr d^3 \bmath k_1^{\rm s}}{(2\pi)^3}\tilde\delta_{\mr g}^{\mr s (1)}(\bmath k_1^{\rm s})\tilde\epsilon^{\mr s (1)}(\hat{\bmath n}|\bmath k^{\rm s}-\bmath k_1^{\rm s}).
\label{eq:d_obs}
\eeq
In the following we calculate the impact of an orientation dependent selection function on the galaxy bispectrum by introducing models for the anisotropic galaxy selection function which are based on symmetry considerations and motivated by the toy models of tidal alignment discussed in Sect.~\ref{sec:TA}. First we extend the linear alignment model from \citet{Hirata09} to second order in the density field, and then construct a new model the anisotropic galaxy selection function due to quadratic alignment.
\subsection{Linear alignment}

In this subsection we construct a model for the anisotropic galaxy selection function $\epsilon$ based on the assumptions that the large scale tidal 
fields induce a preferred direction in galaxy formation, and that the alignment is of linear order in the tidal field. Additionally we require the 
average of $\epsilon(\hat{\bmath n}|\bmath x)$ over the sky to vanish. Then the only possible contraction of the tidal field with the viewing 
direction $\hat{\bmath n}$ is
\beq
\nonumber \epsilon(\hat{\bmath n}|\bmath x)
\!\!&=&\!\!
\frac{A_1}{4\pi G a^2\bar{\rho}_{\mr m}(a)}\left(\hat{n}_i \hat{n}_j\nabla_i\nabla_j-\frac{1}{3}\nabla^2\right)\Psi(\bmath x)\\
&=&\!\!
A_1\hat{n}_i \hat{n}_j\left(\nabla_i\nabla_j\nabla^{-2}-\frac{1}{3}\delta_{ij}\right)\delta(\bmath x),
\eeq
where $\Psi$ is the Newtonian potential, $a$ is the scale factor, and where we have used the Poisson equation to write $\epsilon$ in terms of the 
dimensionless tidal field. $A_1$ is an expansion coefficient which encodes the degree to which galaxy orientations are non-random due to tidal fields 
and the strength of galaxy orientation-dependent selection effects. Note that {\em both} effects need to be present in order to have $A_1 \ne 0$.

To second order in the linear matter density field the anisotropic selection function in Fourier space can be written as
\be
\tilde\epsilon(\hat{\bmath n}|\bmath k)\approx A_1\left[\left(\hat{\bmath n}\cdot \hat{\bmath 
k}\right)^2-\frac{1}{3}\right]\left[\tilde\delta^{(1)}(\bmath k)+\tilde\delta^{(2)}(\bmath k)\right].
\ee
This expression is transformed to redshift space by Taylor expanding the real space expression and using Eqs. (\ref{eq:s}, \ref{eq:f})
\beq
\nonumber \epsilon^{\rm s}(\hat{ \bmath n} |\bmath x^{\rm s}) \!\!&=&\!\! \epsilon(\hat{ \bmath n} |\bmath x)  \approx \epsilon(\hat{ \bmath n}|\bmath x ^{\rm s}) + \left(\bmath x - \bmath x^{\rm s}\right)\cdot \nabla  \epsilon(\hat{ \bmath n}|\bmath x ^{\rm s}) +\mathcal{O}(\delta^3)\\
&=& \!\! \epsilon(\hat{ \bmath n}|\bmath x ^{\rm s})  + f \hat{\bmath n}\cdot \nabla\; \nabla^{-2}\delta^{(1)}(\bmath x^{\rm s})\; \hat{\bmath n}\cdot \nabla \epsilon (\hat{ \bmath n} |\bmath x^{\rm s})\;,
\eeq
and hence in Fourier space
\beq
\nonumber \tilde \epsilon^{\rm s(1)}(\hat{ \bmath n}| \bmath k ^{\rm s}) \!\!\!\!& = &\!\!\!\! \tilde \epsilon^{\rm (1)}(\hat{ \bmath n}| \bmath k ^{\rm s})\\
\nonumber \tilde \epsilon^{\rm s(2)}(\hat{ \bmath n}| \bmath k ^{\rm s}) \!\!\!\! & = &\!\!\!\! \tilde \epsilon^{\rm (2)}(\hat{ \bmath n}| \bmath k^{\rm s})\\
& + &\!\!\!\!\!\!\! 
\int\!\! \frac{\rm d^3 \bmath k_1^{\rm s}}{(2 \pi)^3}
f \mu_1 \mu_{\bmath k^{\rm s}-\bmath k_1^{\rm s}} \frac{k_1^{\rm s}}{|\bmath k^{\rm s}-\bmath k_1^{\rm s}|}
\tilde\delta^{(1)}(\bmath k^{\rm s}-\bmath k_1^{\rm s}) \tilde \epsilon^{(1)}(\hat{\bmath n} | \bmath k_1^{\rm s}).
\eeq
Using this form for the selection function in combination with Eq.~(\ref{eq:d_obs}), we now calculate the galaxy bispectrum modulated by linear tidal alignments. Then the first order observed density contrast is given by
\be
\tilde \delta_{\rm g}^{\rm{obs}(1)}(\mathbf k^{\rm s}) = 
\tilde \delta^{(1)}(\mathbf k) \left(b_1-\frac{1}{3}A_1 + (A_1 + f)\mu_1^2\right)\;.
\ee
The different terms contributing to the observed galaxy bispectrum can be calculated as
\beq
\nonumber \ensav{\tilde \delta_{\rm g}^{\rm{obs}(1)}(\mathbf k_1^{\rm s})\tilde \delta_{\rm g}^{\rm{obs}(1)}(\mathbf k_2^{\rm s})\;\tilde \delta_{\rm g}^{\rm s(2)}(\mathbf k_3^{\rm s})} \!\!\!\!\! &=  &\!\!\!\!\!
(2\pi)^3 \delta_{\mr D}\left(\bmath k^{\rm s}_{123}\right)P(k_1^{\rm s})P(k_2^{\rm s})\\
\nonumber&\times &\!\!\!\!\!  \left(b_1-\frac{1}{3}A_1 + (A_1 + f)\mu_1^2\right)\\
\nonumber &\times &\!\!\!\!\!  \left(b_1-\frac{1}{3}A_1 + (A_1 + f)\mu_2^2\right)  \\
&\times &\!\!\!\!\! 2\; Z_2(\mathbf k_1^{\rm s},\mathbf k_2^{\rm s})\;,\\
\nonumber \ensav{\tilde \delta_{\rm g}^{\rm{obs}(1)}(\mathbf k_1^{\rm s})\tilde \delta_{\rm g}^{\rm{obs}(1)}(\mathbf k_2^{\rm s})\;\tilde \epsilon^{\rm s(2)}(\hat{\mathbf n}|\mathbf k_3^{\rm s})} \!\!\!\!\!& = &\!\!\!\!\!
(2\pi)^3 \delta_{\mr D}\left(\bmath k^{\rm s}_{123}\right)  P(k_1^{\rm s})P(k_2^{\rm s}) \\
\nonumber&\times &\!\!\!\!\!  \left(b_1-\frac{1}{3}A_1 + (A_1 + f)\mu_1^2\right)\\
\nonumber &\times &\!\!\!\!\!  \left(b_1-\frac{1}{3}A_1 + (A_1 + f)\mu_2^2\right)\\
\nonumber &\times &\!\!\!\!\! \left\{2 A_1 \left(\mu_{12}^2-\frac{1}{3} \right)F_2(\bmath k_1, \bmath k_2)\right.\\
\nonumber & + & \!\!\!\!\!A_1f \mu_1\mu_2\frac{k_1^{\rm s}}{k_2^{\rm s}}\left(\mu_{1}^2-\frac{1}{3} \right)\\
&+ &\left. \!\!\!\!\!A_1f \mu_1\mu_2\frac{k_2^{\rm s}}{k_1^{\rm s}}\left(\mu_{2}^2-\frac{1}{3} \right)\right\}\,
\eeq
and the contribution from the last term in Eq. (\ref{eq:d_obs}) containing a convolution of first order density contrast and anisotropic selection function
\beq
\nonumber \ensav{\tilde \delta_{\rm g}^{\rm{obs}(1)}(\mathbf k_1^{\rm s})\tilde \delta_{\rm g}^{\rm{obs}(1)}(\mathbf k_2^{\rm s})\;\left(\tilde \delta_{\rm g}^{\rm s(1)}\otimes\tilde \epsilon^{\rm s(1)}\right)(\hat{\mathbf n}, \mathbf k_3^{\rm s})}  = 
(2\pi)^3 \delta_{\mr D}\left(\bmath k^{\rm s}_{123}\right)P(k_1^{\rm s})P(k_2^{\rm s})&\\
\nonumber\times \left(b_1-\frac{1}{3}A_1 + (A_1 + f)\mu_1^2\right)\left(b_1-\frac{1}{3}A_1 + (A_1 + f)\mu_2^2\right)  & \\
\times\; A_1 \left\{\left(b_1 +f \mu_1^2\right)\left(\mu_2 -\frac{1}{3}\right)+ \left(b_1 +f \mu_2^2\right)\left(\mu_1 -\frac{1}{3}\right)\right\}. &
\eeq
Hence the galaxy bispectrum modulated by linear tidal alignments is given by
\beq
B_{\mr g}^{\mr{s, LA}}(\bmath k_1^{\rm s},\bmath k_2^{\rm s},\bmath k_3^{\rm s}) \!\!\!\! &=& \!\!\!\!
\left[b_1-\frac{A_1}{3}+ (A_1+f) \mu_1^2\right]
\nonumber \\ && \!\!\!\! \times \left[b_1-\frac{A_1}{3}+ (A_1+f) \mu_2^2 \right]
\nonumber \\ && \!\!\!\! 
\times \Biggl\{ 2 Z_2(\bmath k_1^{\rm s}, \bmath k_2^{\rm s})
+ 2 A_1\left(\mu_{12}^2-\frac{1}{3} \right)F_2(\bmath k_1^{\rm s}, \bmath k_2^{\rm s})
\nonumber \\ && \!\! \!\!
+ A_1\left[ b_1 \left(\mu_1^2+\mu_2^2-\frac{2}{3}\right)+\frac f3 \left(6 \mu_1^2 \mu_2^2-\mu_1^2-\mu_2^2\right)\right]
\nonumber \\ && \!\! \!\!
+A_1f \mu_1\mu_2\left[\frac{k_2^{\rm s}}{k_1^{\rm s}}\left(\mu_{2}^2-\frac{1}{3} \right) + \frac{k_1^{\rm s}}{k_2^{\rm s}}\left(\mu_{1}^2-\frac{1}{3} \right)\right]
\nonumber \\ && \!\! \!\!
\Biggr\}P(k_1^{\rm s})P(k_2^{\rm s})+\mr{2~perm.}
\label{eq:B_LA}
\eeq

\subsubsection{Transverse galaxy bispectrum}

As the full redshift space bispectrum is a complicated function of configurations described by 5 parameters (3 parameters specifying triangle 
shape, and 2 angles describing the orientation with respect to the line of sight), we will now simplify Eq.~(\ref{eq:B_LA}) by considering only 
triangles 
in the plane of the sky ($\mu_i = 0$), which are the easiest to model and are the triangles observed in photometric redshift surveys.  In this case, 
we find a galaxy bispectrum
\beq
\nonumber B_{\mr g}^{\mr{LA,\perp}}(\bmath k_{1},\bmath k_{2},\bmath k_{3}) = \left(b_1 - \frac{A_1}{3}\right)^2\left[2\left(b_1 - \frac{A_1}{3}\right) 
F_2\left(\bmath k_{1},\bmath k_{2}\right)\right.\\
\left.+b_2-\frac{2}{3}A_1 b_1\right]P(k_1)P(k_2) + \mr{2~perm}.
\label{eq:D_BLA}
\eeq
Comparing this expression to Eq.~(\ref{eq:B_g}), one finds that the effect of linear tidal alignments on the transverse galaxy bispectrum can be 
described as a rescaling of the galaxy bias parameters
\be
b_1 \rightarrow b_1 - \frac{A_1}{3},\;\;\; b_2 \rightarrow b_2 - \frac{2}{3}A_1 b_1.
\label{eq:bias_LA}
\ee
\citet{Hirata09} found that the same rescaling of $b_1$ applies to the real-space ($\mu_i=0$) galaxy power spectrum.  Therefore, {\em the use of the 
real-space power 
spectrum and bispectrum to eliminate galaxy bias parameters and extract $\sigma_8$ is robust against linear tidal alignments}.  However, this 
robustness does not extend to the $\mu_i\neq0$ modes.

For later use, we also write out the systematic error in the transverse galaxy bispectrum induced by linear alignment
\beq
\nonumber \Delta B_{\mr g}^{\rm{LA},\perp}(\bmath k_{1},\bmath k_{2},\bmath k_{3}) \!\!\!&=&\!\!\!\Biggl[2\left(b_1^2A_1 -b_1 \frac{A_1^2}{3} 
+\frac{A_1^3}{27}\right) F_2\left(\bmath k_{1},\bmath k_{2}\right) \\
&&\!\!\! -b_1\frac{A_1}{3}\Biggr] P(k_1)P(k_2)\;+ \mr{2~perm}.
\eeq

\subsubsection{Normalization}
\label{sec:normLA}

Following \citet{Hirata09},  we use $A_1 \approx -0.024$ for LRG-type elliptical galaxies. This is a rough estimate which is based on the assumption that elliptical galaxies are optically thin triaxial systems, that the deviation from spherical symmetry can on average be related to the tidal field (with correlation strength $B$), on different models for the orientation dependence of a galaxy's apparent magnitude (parametrized by $\chi$), and the slope of the galaxy luminosity function $\eta$:
\be
A_1 = 2 \eta \chi B\;.
\label{eq:A1}
\ee
While the total flux of an optically thin galaxy is not affected by tidal alignments, the average isophotal ellipticity and projected effective radius of a galaxy become a function of the tidal field.

The selection of galaxies in a survey will be modified by tidal alignment if part of the selection criteria is a magnitude cut, and if the apparent magnitude of a galaxy depends on its orientation. The apparent magnitude of a galaxy is nearly orientation independent if measured using Petrosian magnitudes or model magnitudes which are based on an accurate model for the radial profile, then at the level of the toy model considered by \citet{Hirata09} $\epsilon \approx 0$.

If galaxies are selected using isophotal magnitudes or aperture magnitudes, more light will be counted if a galaxy is viewed along its long axis than its short axis. The selection factor $\chi$ in 
Eq.~(\ref{eq:A1}) depends on the method used to measure galaxy fluxes \citep[c.f. Fig. 2 in][]{Hirata09}, and it translates the fractional change in effective radius induced by intrinsic alignment to a 
fractional change in measured flux. This change in measured flux moves galaxies across the selection threshold, and it is translated into change in number density by assuming a luminosity function with slope $-\eta$.

The strength of the tidal alignment effect $B$ is determined from measurements of the density-ellipticity cross-correlation function \citep{H07}. Our chosen normalization further assumes a LRG luminosity function with $\eta = 4.0$ and galaxy selection based on isophotal magnitudes measured within $\sim3$ effective radii. Also note that this normalization is based on observations around $z=0.3$ and should only be used near this redshift as the LRG luminosity function and the correlation between tidal field and galaxy orientation may show strong evolution with redshift.

\subsection{Quadratic alignment}

The leading-order alignment of galactic angular momentum in tidal torque theories is quadratic in the tidal tensor because of the need for both a tidal 
field and an anisotropic inertia tensor on which it can act.

The anisotropic selection function for a disc galaxy is generally a function of its inclination $i$ (defined by $\cos i=\hat{\bmath L}\cdot\hat{\bmath 
n}$).  While $i$ is in the range $0\le i\le\pi$, we expect most selection criteria to be symmetric with respect to an observer being above or below the 
plane of the target, so it follows that the anisotropic part of the selection function contains only even-order spherical harmonics:
\be
\Upsilon({\mathbfss Q}\hat{\bmath n},\bmath x) = \sum_{J\ge2,\,{\rm even}} c_{J} P_{J}(\cos i), 
\ee
where $P_J$ is a Legendre polynomial.  
Using Eq.~(\ref{eq:defepsilon}),
and noting that
for a disk galaxy, we may replace the general integration over orientations ${\mathbfss Q}\in$SO(3) with an integration over directions of the angular 
momentum vector $\hat{\bmath L}\in S^2$, we may write
\be
\epsilon(\hat{\bmath n}|\bmath x)
= \sum_{J\ge2,\,{\rm even}} c_J
 \int_{S^2} p(\hat{\bmath L}|{\bmath x}) P_J(\cos i) \,{\rm d}^2\hat{\bmath L}.
\label{eq:ee}
\ee
Because the quadratic alignment
model contains two factors of the tidal field, which are spin 2, $p(\hat{\bmath L}|{\bmath x})$ can contain 
spherical harmonics only through order $J\le4$.  For simplicity, we will focus only on the quadrupolar $J=2$ term in the sum (while noting that the 
hexadecapolar alignment $J=4$ is in principle possible).  Then Eq.~(\ref{eq:ee}) implies that
\be
\epsilon(\hat{\bmath n}|\bmath x)\propto \langle P_2(\hat{\bmath L}\cdot\hat{\bmath n}) \rangle,
\ee
where the average is taken over the local probability distribution of $\hat{\bmath L}$.
Equivalently, using Eq.~(\ref{eq:LiLj}), we find that
\be
\epsilon(\hat{\bmath n}|\bmath x) = \tilde A_2
\left( \hat n_i \hat n_j - \frac13\delta_{ij}\right) \hat T_{ih}\hat T_{hj}.
\label{eq:e2that}
\ee
We relate $\hat T_{ij}$ to the dimensionless shear field tensor $T_{ij}$,
\beq
\nonumber \tilde T_{ij}(\bmath k) \!\!\! &=& \!\!\! \frac{1}{4\pi G a^2\bar{\rho}_{\mr m}(a)}\left(k_i k_j - \frac{1}{3}\delta_{ij}k^2\right) 
\tilde\Psi(\bmath k) \\
&=& \!\!\! \left(\hat k_i \hat k_j - \frac{1}{3}\delta_{ij}\right) \tilde\delta(\bmath k),
\label{eq:that}
\eeq
by approximating the scalar $T^2\equiv T_{ij}T_{ji}$ with its expected value $C^2$:
\beq
C^2 \equiv \langle T^2\rangle = \frac23\sigma^2(R),
\label{eq:C2}
\eeq
i.e. we approximate $\hat T_{ij}\approx C^{-1}T_{ij}$.
As this expression for the anisotropic selection function is already second order in the density field, effects associated with mapping $\epsilon$ to redshift space only enter at higher orders than considered in this analysis and in the following we will drop the superscript ${\rm s}$ to denote Fourier modes in redshift space.

Note that $C^2$ is proportional to the variance of the smoothed density field smoothed on the halo collapse scale $R$,
since the density and tidal fields are both derived by taking second 
derivatives of the potential.

Then the contribution of quadratic alignment to the orientation dependent selection function can be written as
\beq
\nonumber \tilde\epsilon^{(2)}(\hat{\bmath n}|\bmath k) \!\!\! &=& \!\!\!
\tilde A_2 \left( \hat n_i \hat n_j -\frac{1}{3}\delta_{ij}\right) \int\frac{\mr d^3 \bmath k'}{(2\pi)^3}\hat{\tilde{T}}_{ih}(\bmath 
k)\hat{\tilde{T}}_{hj}(\bmath k'')\\
\nonumber  &=& \!\!\! A_2\hat n_i \hat n_j 
\int\frac{\mr d^3 \bmath k'}{(2\pi)^3}
\left\{
\left(\hat k'_i \hat k'_h - \frac{1}{3}\delta_{ih}\right)
\left(\hat k''_h \hat k''_j - \frac{1}{3}\delta_{hj}\right)
\right.\\
&& \left.
-\frac{1}{3}\delta_{ij}
\left[\left( \hat{\bmath k}'\cdot\hat{\bmath k}''\right)^2-\frac{1}{3}\right]\right\}
 \tilde\delta^{(1)}(\bmath k') \tilde\delta^{(1)}(\bmath k''),
\label{eq:e2t}
\eeq
where $\bmath k'' = \bmath k - \bmath k'$.
This term contributes to the observed galaxy bispectrum via
\beq
\Delta B_{\mr g}^{\mr{QA}}(\bmath k_1,\bmath k_2,\bmath k_3) \!\!\!\!\! &=& \!\!\!\!\!
2A_2
\left(b_1-\frac{A_1}{3}+ (A_1+f) \mu_1^2\right)
\nonumber \\
&& \!\!\!\!\! \times
 \left(b_1-\frac{A_1}{3}+ (A_1+f) \mu_2^2\right)P(k_1)P(k_2)
\nonumber \\
&& \!\!\!\!\!
\times\left\{\mu_1\mu_2\hat{\bmath k}_1\cdot\hat{\bmath k}_2-
\frac{1}{3}
\left( \mu_1^2+\mu_2^2+(\hat{\bmath k}_1\cdot\hat{\bmath k}_2)^2  \right)+\frac{2}{9}\right\}
\nonumber \\
&& \!\!\!\!\!
+\;\mr{2~perm}.
\eeq
Here $A_1\neq 0$ if the galaxy population under consideration is also subject to linear alignment, and we have defined $A_2\equiv\tilde A_2/C^2$.

\subsubsection{Transverse galaxy bispectrum}

\begin{figure}
\includegraphics[width = 3.2in]{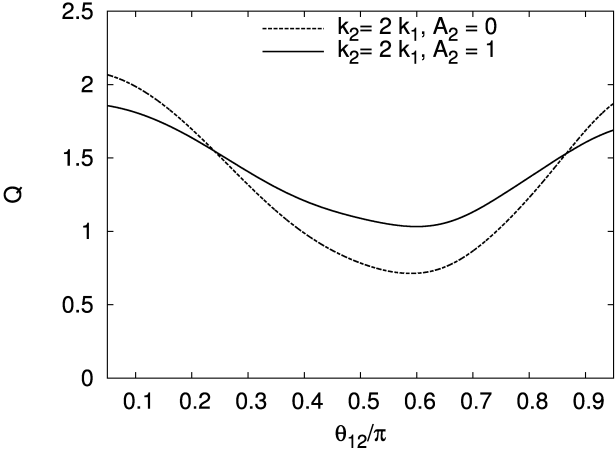}
\caption{Effect of quadratic alignment on the reduced transverse galaxy bispectrum with $b_1 = 1$, $k_1 = 0.05 h\,$Mpc$^{-1}$, and where $\theta_{12}$ denotes the 
angle between $\bmath k_1$ and $\bmath k_2$, and for $A_2=1$.}
\label{fig:Q}
\end{figure}

\begin{figure}
\includegraphics[width = 3.2in]{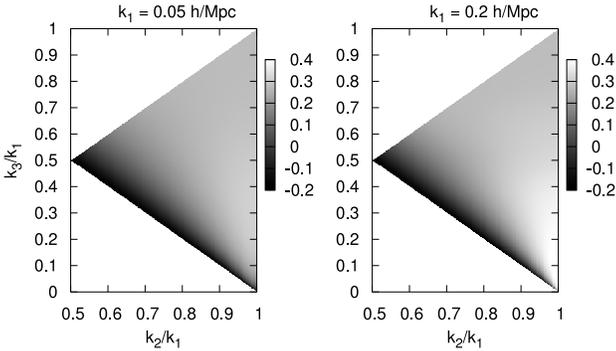}
\caption{Systematic offset of the reduced transverse galaxy bispectrum due to quadratic alignment with $b_1 = 1$ and $A_2=1$ as a function of triangle shape and scale. Shown are all possible closed triangle configurations with $k_1\ge k_2\ge k3$ for a given $k_1$, areas in configuration space which do not correspond to a closed triangle are shown in white (located around the top and bottom left corner of each plot).  Equilateral triangles are located in the upper right corner of the configuration space, isosceles triangles lie on the upper diagonal, and collinear ($\theta_{12}\rightarrow 0$) triangles near the lower diagonal.}
\label{fig:conf}
\end{figure}

The quadratic alignment model modifies the observed transverse galaxy bispectrum by
\beq
\Delta B_{\mr g}^{\mr{QA,\perp}}(\bmath k_{1},\bmath k_{2},\bmath k_{3}) \!\!\!\! &=& \!\!\!\! \frac23A_2b_1^2\left[\frac{2}{3}-\left(\hat{\bmath 
k}_1\cdot\hat{\bmath k}_2  \right)^2\right]P(k_1)P(k_2)
\nonumber \\ && \!\!\!\!
+\;\mr{2~perm.}
\label{eq:B_QA}
\eeq
Note that this systematic offset is independent of $b_2$, and its amplitude scales linearly with $A_2$ and quadratically with $b_1$. The systematic offset cannot be expressed as a simple rescaling of the galaxy bias parameters due to its shape dependence. Figure~\ref{fig:Q} illustrates its effect on the reduced transverse galaxy bispectrum
\be
Q_{\mr g}(\bmath k_1,\bmath k_2,\bmath k_3) = \frac{B_{\mr g}(\bmath k_1,\bmath k_2,\bmath k_3)}{P_{\mr g}(k_1)P_{\mr g}(k_2)+P_{\mr g}(k_1)P_{\mr g}(k_3) + P_{\mr g}(k_2)P_{\mr g}(k_3)}\;,
\ee
which is only mildly dependent on cosmology as the amplitude of fluctuations has been divided out. The shape and scale dependence of $\Delta Q_{\mr g}$ is further illustrated in Fig.~\ref{fig:conf}, which shows the systematic offset for all possible closed triangle configurations with $k_1\ge k_2\ge k3$, with the left plot showing triangles with $k_1 = 0.05 h/{\mr{Mpc}}$ and the right plot showing triangles with $k_1 = 0.2 h/{\mr{Mpc}}$. The systematic offset is negative for triangles which are close to collinear, and for the scales considered in this analysis it shows little scale dependence.


\subsubsection{Normalization}
\label{sec:normQA}

Similar to the normalization of the linear alignment contamination outlined in Section~\ref{sec:normLA}, the magnitude of the observed contamination 
due to quadratic alignment again depends on (i) the orientation dependence of the recovered flux (continuum or line), (ii) the slope of the galaxy 
luminosity function, and (iii) the strength of the tidal alignment effect.  We may use models for (i) and direct measurements for (ii), but (iii) is 
harder.  For the linear alignment model we were able to use the observational constraints from the density-ellipticity cross-correlation function, but 
this is not an option here as the quadratic alignment contribution to two point statistics vanish to leading order.  Another option would be to set 
limits using the observed ellipticity variance, which must set an upper limit on $\alpha^2$ (this was the approach followed in \citealt{CNPT01} for 
estimating the intrinsic ellipticity correlation contamination of weak lensing surveys).  We will take an even simpler approach here, and use some 
simple theoretical arguments on the value of $\alpha$.

In the tidal torque model, the distribution of disk normal vectors $\hat{\bmath L}$ given some tidal tensor $\hat{\mathbfss T}$ can be approximated by
\citep{CNPT01}
\be
p(\hat{\bmath L}|\hat{\mathbfss T}) \approx\frac{1}{4\pi}
\left(1+\frac{3\alpha}{2}-\frac{9\alpha}{2}\hat L_{i}\hat L_{j}\hat T_{ik}\hat T_{jk}\right).
\label{eq:pLT}
\ee
For a geometrically thin disk with normal vector $\hat{\bmath L}$ observed along the $\hat z$ axis, the inclination is $\cos i = \hat L_3$.  The 
following constraints can be placed on $\alpha$:
\begin{itemize}
\item Since $\hat L_{i}\hat L_{j}\hat T_{ik}\hat T_{jk}$ can take on any value between $0$ and $\frac23$,
the requirement that $p(\hat{\bmath L}|\hat{\mathbfss T})\ge 0$ sets the constraint $|\alpha|\le\frac23$.
\item If one neglects correlations between the external tidal field and the moment of inertia tensor of the collapsing protogalaxy, one finds 
$\alpha=\frac35$ \citep{LP00}.
\item The angular momentum of the disc of a galaxy may be disaligned from that of its host halo, due to e.g. torques between the disc and halo, or due 
to the disc containing only a specially selected subset of the halo's baryons.  For a Gaussian distribution of disalingment angles with rms 
per axis $\Theta$, the 
$JM$ spherical harmonic component of $p(\hat{\bmath L}|\hat{\mathbfss T})$ is suppressed by a factor of $\exp[-J(J+1)\Theta^2/2]$; since we have a 
quadrupolar anisotropy ($J=2$), $\alpha$ is suppressed by a factor of $\exp(-3\Theta^2)$.
\end{itemize}
The above arguments suggest that $|\alpha|$ of several tenths is plausible, but in no case should it exceed $\frac23$.  Also, while the simplest 
version of the tidal torque hypothesis implies $\alpha>0$, there is no physical reason why negative values should not be allowed.

Next we determine the relation between an inclination dependent observed flux and the selection function $\epsilon$: Assume a galaxy flux distribution 
with slope $\mr d \ln \bar n /\mr d \ln F_{\mr min} = -\eta$. Then the number density of galaxies per logarithmic range in the intrinsic flux $F_{\mr 
i}$ per unit solid angle of disk orientation is 
\be
\mathcal N (F_{\mr i}, \hat{\bmath L})\propto F_{\mr i}^{-\eta}{p}(\hat{\bmath L}|T)\;.
\ee
Let the observed, inclination dependent flux be $F(i) =F_{\mr i}\Phi(i)$. The number density of galaxies above some threshold flux $F_0$ then evaluates 
to
\beq
\mathcal N(> F_0) \!\!\!\! &\propto& \!\!\!\! \int \mr d^2 \hat{\bmath L}\int_{F_0/\Phi(i)}^{\infty}\mr d\ln F_{\mr i}F_{\mr 
i}^{-\eta}{p}(\hat{\bmath L}|\hat {\mathbfss T})
\nonumber \\ &\propto& \!\!\!\!
\int_0^{\pi}\left[\Phi(i)\right]^\eta \left[1-\frac{9\alpha}{2}\left(\hat T_{3j}^2-\frac{1}{3}\right)P_2(\cos(i))\right]\sin i\;\mr d i,
\eeq
where we have performed both the integral over $\phi$ and over $F_{\rm i}$ (since the latter is simply a power law), and defined $\hat T_{3j}^2\equiv 
\hat T_{3j}\hat T_{3j}$.
Defining
\be
\psi =\frac{\int_0^{\pi}\left[\Phi(i)\right]^{\eta}P_2(\cos i)\sin i\;\mr di}{\int_0^{\pi}\left[\Phi(i)\right]^{\eta}\sin i \;\mr di},
\label{eq:psi}
\ee
the anisotropic part of the observed galaxy count can be written as
\be
\epsilon(\hat z|\bmath x) = -\frac{9\alpha}{2}\psi\left(\hat T_{3j}^2-\frac{1}{3}\right).
\ee
Combining this with Eq.~(\ref{eq:e2that}), we conclude that
$\tilde A_2 = -\frac92\alpha\psi$, and hence
\be
A_2 = -\frac92\frac\alpha{C^2}\psi = -\frac{27}4\frac{\alpha\psi}{\sigma^2_\delta(R)}.
\label{eq:A2PsiK}
\ee
The top-hat variance is related to the bias of the galaxies if the mass function is nearly universal \citep{ShethTormen}; for example, at $b=1$ we have 
$\sigma^2_\delta(R)=2.96$, whereas at $b=2$ we have $\sigma^2_\delta(R)=0.83$.

The last step in obtaining a numerical estimate for $A_2$ is evaluating the orientation dependent selection factor $\psi$. 
This requires a model for the angular distribution of emitted radiance $\Phi(i)$, which also determines the selection 
probability $p(i) \propto \left[\Phi(i)\right]^\eta$. Several geometric toy models for the vertical distributions of emitters and dust are discussed 
by \citet{Hirata09}, and for galaxy distributions with $\eta \approx 2$ (appropriate for [O{\sc\,ii}] and H$\alpha$ surveys), $\psi$ is found 
to be of order a few tenths: for example, it is $\psi=0.4$ in the optically thick slab model; $\psi=0.23$ (0.30) in the uniform slab model with normal 
optical depth $\tau=0.5$ (1.0); and $\psi=0.26$ (0.37) in the sheet-in-slab model with $\tau=0.5$ (1.0).

These toy models suggest that $A_2$ will be of order unity and we assume $A_2 =1$ for illustrative purpose in the following 
analysis.\footnote{In principle, either {\em sign} of $A_2$ is allowed by our above calculations; for negative $A_2$ the direction of the parameter 
biases should be reversed.}  For application 
to any survey the normalization must be calculated based on the detailed selection criteria and galaxy distribution.

\section{Fisher Matrix Analysis}
\label{sec:fisher}

We now estimate the parameter bias induced by a tidal alignment contamination by performing a Fisher matrix analysis for a survey with characteristics 
similar to the Dark Energy Survey (DES)\footnote{URL: \tt http://www.darkenergysurvey.org/}, assuming that one would use the angular bispectrum of a 
slice of galaxies in photometric redshift space.  A spectroscopic survey covering a similar volume and oversampling the density field ($nP>1$) would of 
course yield tighter constraints, but a full Fisher analysis of such a survey including redshift space distortions and finger-of-God parameters is 
beyond the scope of this paper. 

\subsection{Survey characteristics and analysis details}

Our fictitious survey has the same area as the DES, $\Omega = 5000$ square degrees. We assume a constant comoving galaxy density over the redshift 
range of interest and use a radial galaxy selection function of the form expected for the DES \citep{NPR10},
\be
\frac{\rm d\,Prob}{dz} \propto \left(\frac{z}{0.5}\right)^2 \exp\left(-\frac{z}{0.5}\right)^{1.5}\;,
\label{eq:phi}
\ee
In order to project out redshift space distortions we consider the angular clustering of galaxies projected over a finite radial distance. For our theoretical modeling the projection over a finite range in radial distance is equivalent to a projection over a finite redshift range, and we choose $0.4 \leq z \leq 0.6$. Observationally, this mapping is complicated by the distribution of photometric redshifts and the effect of redshift space distortions on the boundary of a region selected in redshift space \citep[e.g.][]{PSS07,NPR10}.

\subsubsection{Binned angular multispectra and covariances}

We calculate the angular power and multispectra $\mathcal P_{N}$ using the Limber equation in Fourier space \citep{Kaiser92, BKJ00}:
\be
\mathcal P_{N}\left(\bmath l_1\ldots \bmath l_N\right) =\int_{z=0.4}^{z=0.6}\mr d \chi\; \frac{\phi^N(\chi)}{\chi^{2N-2}} P_{N}\left(\frac{\bmath l_1}{\chi},\ldots,\frac{\bmath l_N}{\chi};\chi\right),
\label{eq:Limber}
\ee
where $P_{N}$ is the three dimensional $N$-point correlation function in Fourier space. In the following we use $\mathcal P,\; \mathcal B,\; \mathcal 
T$ to denote the angular \emph{galaxy} power spectrum, bispectrum and trispectrum.

For a linear alignment contamination, the change in the observed angular galaxy bispectrum is described by the same bias parameter rescaling (Eq.~(\ref{eq:bias_LA})) as for the transverse galaxy bispectrum discussed above.
The magnitude of the systematic offset in the angular galaxy bispectrum induced by a quadratic alignment contamination is proportional to $A_2 b_1^2$ and independent of $b_2$. As the angular projection mixes different physical scales, the exact configuration dependence and normalization of the angular bispectrum contamination depends strongly on the radial selection function \citep[for details see][]{FT99}. As can be seen from Fig.~\ref{fig:conf} the systematic offset on the reduced transverse galaxy bispectrum is only weakly scale dependent, thus with our choice for the radial selection funtcion the angular reduced bispectrum has very similar shape dependence.

The Limber approximation requires the transverse scales under consideration to be significantly smaller than the radial projection depth, hence we 
limit our analysis to angular scales corresponding to comoving Fourier modes $k \geq 0.04 h\,$Mpc$^{-1}$. As our intrinsic alignment toy models and 
biasing approximation are not designed to describe in the non-linear regime of structure formation, we will only consider angular frequencies corresponding to 
\be
0.04\;  h{\,\rm Mpc}^{-1} \leq k \leq 0.2\;  h{\,\rm Mpc}^{-1}.
\label{eq:scales}
\ee 
We approximate the galaxy power spectrum by the linear matter power spectrum rescaled by the linear bias (Eq.~\ref{eq:P_g}); bispectra and trispectra 
on these scales 
are 
approximated by the tree-level perturbation theory in combination with local biasing (Eq.~\ref{eq:biasing}), i.e. using Eqs.~(\ref{eq:P_g}), 
(\ref{eq:B_g}), and
(\ref{eq:T_g}). These are evaluated using transfer functions generated by CMBFAST \citep{CMBFAST} for the best-fit WMAP 7 cosmology \citep{WMAP7}. 
Compared to an approach combining the halo model with halo occupation distribution modeling \citep[e.g.][]{BW02,CS02} this is computationally much 
faster, the only model input is our biasing prescription and does not require halo models for intrinsic alignment. In the large scale limits the halo models asymptote to the perturbation theory result, and 
at the scales of our analysis the galaxy power spectrum is fairly well described by perturbation theory \citep{C04, S+08}. At redshift $z= 0$, 
\citet{S+08} find the reduced halo model bispectrum with $k_2 = 2 k_1$ to be in close agreement with perturbation theory results at scales $k_1\leq 
0.1\; h/{\mr {Mpc}}$,  except for collinear configuration ($\theta_{12}\rightarrow 0$). As we only consider triangle configurations with all angular 
frequencies $k_{1,2,3} \leq 0.2\;  h\,$Mpc$^{-1}$, the perturbation theory results should be sufficient at the level of this analysis. However, at scales smaller than $k \sim 0.1 h/\mr{Mpc}$ \citet{S+08} and \citet{GJ09B} find the bispectrum measured from simulations to differ at the 10-20\% level from the perturbation theory. Note that these systematic effects on the determination of bias parameters on small scales are larger than the tidal alignment contaminations discussed here.

We model the observed power spectrum by averaging the angular power spectrum over bins of width $\Delta l$,
\be
\mathcal P(\bar l) \equiv \int_{\bar l -1/2\Delta l}^{\bar l +1/2\Delta l}\frac{\mr d l\; l}{\bar l \Delta l}\mathcal P(l),
\ee
and the corresponding covariance is given by
\beq
\nonumber \mr{Cov}\left(\mathcal P(\bar l_1) \mathcal P(\bar l_2)\right)
\!\!\!\! &=& \!\!\!\! \frac{1}{\Omega}\left\{\delta_{\bar l_1,\bar l_2}\frac{4 \pi}{\bar l_1 
\Delta l}\left[\mathcal P(\bar l_1)+\frac{1}{\bar n}\right]^2\right.\\
\\
&& +\left.\int_1\int_2\mathcal T(\bmath l_1,-\bmath l_1,\bmath l_2,-\bmath l_2)\right\}\;,
\label{eq:CovP}
\eeq
where $\bar n$ is the average projected density of the galaxy population under consideration. Here the first term is a combination of Gaussian cosmic 
variance and shot noise. The second term involving the trispectrum of parallelogram configurations  is the non-Gaussian power spectrum covariance.

The bispectrum is sampled with uniform binning $\Delta l$ in all angular frequencies. Defining
\be
\int_i \equiv \int_{\bar l_i -1/2\Delta l}^{\bar l_i 
+1/2\Delta l}\frac{\mr d l_i\; l_i}{\bar l_i \Delta l},
\ee
the bin-averaged bispectrum is given by 
\be
\mathcal B(\bar l_1,\bar l_2,\bar l_3) \equiv \int_1\int_2\int_3 \mathcal B(l_1,l_2,l_3)\delta_{\mr D}(\bmath l_1+\bmath l_2+\bmath l_3).
\ee
We approximate the expression from \citet{JXS10} for the full non-Gaussian covariance of the bin-averaged bispectrum by
\begin{align}
\nonumber & \mr{Cov}\left(\mathcal B(\bar l_1,\bar l_2,\bar l_3)\mathcal B(\bar l_4,\bar l_5,\bar l_6)\right) =
\frac{(2\pi)^3}{\Omega\bar l_1\bar l_2\bar l_3\Delta l^3}\Lambda^{-1}(\bar l_1,\bar l_2,\bar l_3) \\
\nonumber &\;\;\;\;\; \times D_{\bar l_1,\bar l_2,\bar l_3,\bar l_4,\bar l_5, \bar l_6}\left[\mathcal P(\bar l_1)+\frac{1}{\bar n}\right]\left[\mathcal P(\bar l_2)+\frac{1}{\bar n}\right]\left[\mathcal P(\bar l_3)+\frac{1}{\bar n}\right] \\
\nonumber & +\frac{2\pi \Lambda^{-1}(\bar l_1,\bar l_2,\bar l_3)\Lambda^{-1}(\bar l_4,\bar l_5,\bar l_6)}{\Omega}\delta_{\bar l_3,\bar l_4}\int_1\int_2\int_3\int_5\int_6\delta_{\rm D}\left(\bmath l_1+\bmath l_2+\bmath l_3\right)\\
 \nonumber &\;\;\;\;\; \times\Bigg\{\delta_{\rm D}\left(\bmath l_3+\bmath l_5+\bmath l_6\right)\mathcal B(l_1,l_2,l_3)\mathcal B(l_3,l_5,l_6)\\
& \;\;\;\;\; + \delta_{\rm D}\left(-\bmath l_3+\bmath l_5+\bmath l_6\right)\mathcal T(\bmath l_1,\bmath l_2,\bmath l_5,\bmath l_6)\mathcal 
P(l_3)\Bigg\} + 8\mr{\; perm.},
\label{eq:CovB}
\end{align}
where the symmetry factor $D_{\bar l_1 ... \bar l_6}$ is non-zero only for diagonal elements of the covariance ($\{\bar l_1,\bar l_2,\bar l_3\} =\{\bar 
l_4,\bar l_5,\bar 
l_6\}$): $D_{\bar l_1 ... \bar l_6}=1$, 2, or 6 for scalene, isosceles, or equilateral triangles respectively. If $\bar l_1,\bar l_2,\bar l_3$ form a 
triangle, then 
$\Lambda^{-1}(\bar l_1,\bar l_2,\bar l_3)$ is the area of this triangle, otherwise $\Lambda^{-1}$ = 0. The first term is the Gaussian (diagonal) part 
of the covariance which is proportional to the product of three power spectra which have been modified to account for Gaussian shot noise. The second/ 
third terms are non-Gaussian contributions from triangle pairs which have at least one common side so that the pentaspectrum can be factorized into two bispectra/ a trispectrum and a power spectrum. We have dropped a term which is proportional to the general connected pentaspectrum.

\subsection{Biased parameter estimates for galaxy bias parameters}

Having set up a model for the observable data and their covariances, we can now quantify the power of our fictitious survey at constraining model 
parameters using the Fisher matrix
\be
\mathcal F_{\alpha \beta} = \frac{\partial \vec{\mathcal P}^t}{\partial p_{\alpha}} \mr{Cov}^{-1}\left(\vec{\mathcal P}, \vec{\mathcal P}\right)\frac{\partial \vec{\mathcal P}}{\partial p_{\beta}} +  \frac{\partial \vec{\mathcal B}^t}{\partial p_{\alpha}} \mr{Cov}^{-1}\left(\vec{ \mathcal B}, \vec{ \mathcal B}\right)\frac{\partial \vec{\mathcal B}}{\partial p_{\beta}}\;,
\label{eq:fisher}
\ee
where the $\vec{\mathcal P}$ and $\vec{\mathcal B}$ are data vectors with the binned angular \emph{galaxy} power spectrum and bispectrum as data points. The data vectors and their covariances depend explicitly on the bias parameters through Eqs.~(\ref{eq:P_g}, 
\ref{eq:B_g}, \ref{eq:T_g}). Note 
that we do not include cross-correlations between power spectrum and bispectrum, both for simplicity and because they are small in the weakly 
nonlinear regime (but see \citealt{SCPS06} for 
their constraining power in the weakly non-linear regime). The parameters of interest here are the linear and quadratic galaxy bias and we marginalize 
over the normalization of the matter power spectrum $\sigma_8$, i.e. $\bmath{p} = (b_1, b_2,\sigma_8)$. Our fiducial model assumes 
$\sigma_8 =0.8$, no intrinsic alignment contamination, and covers a range of bias parameters, while all other cosmological parameters are fixed to their best-fit WMAP 7 values.

The inverse Fisher matrix serves as a lower limit on the marginalized covariance of statistical parameter errors 
\be
\ensav{\delta p_{\alpha} \delta p_{\beta} } = \left(\mathcal F ^{-1}\right)_{\alpha \beta}\;.
\label{eq:Fisher_error}
\ee
Hence the statistical error on the inferred parameters is inversely proportional to $\sqrt{\Omega}$, as can be seen from the expressions 
(Eqs.~\ref{eq:CovP}, \ref{eq:CovB}) for the data covariances.
The presence of a systematic error $\vec{\Delta \mathcal B}$, $\vec{\Delta \mathcal P}$ in the data which is not included in the model induces a bias in the parameter estimate compared to its fiducial values. To first order it is given by \citep[e.g.][]{H06, AR08}
\begin{align}
\nonumber \Delta p_{\alpha} = \ensav{\hat p_{\alpha}} - p_{\alpha}^{\mr{fid}} =& \left(\mathcal F^{-1}\right)_{\alpha \beta}\; \left[ \vec{\Delta \mathcal P^{t}}\; \mr{Cov}^{-1}\left(\vec{ \mathcal P}, \vec{ \mathcal P}\right)\frac{\partial \vec{\mathcal P}}{\partial p_{\beta}}\right.\\
&+\left. \vec{\Delta \mathcal B^{t}}\; \mr{Cov}^{-1}\left(\vec{ \mathcal B}, \vec{ \mathcal B}\right)\frac{\partial \vec{\mathcal B}}{\partial p_{\beta}}\right]\;,
\label{eq:Fisher_bias}
\end{align}
where the data vectors and covariances are evaluated at the fiducial model.

This systematic bias is independent of the survey area, but it is influenced by our choice of survey parameters through the selection function (Eq.~\ref{eq:phi})
 and data binning scheme. It also depends on projected number density of the galaxy population of interest as $\bar n$ determines the importance of shot noise.
We adopt a uniform sampling with $20$ equidistant bins in all angular frequencies ($l_1,l_2,l_3$) corresponding to Eq.~(\ref{eq:scales}) 
and assume a projected density of $\bar n  = 1/\mr{arcmin}^2$ for a galaxy population in the redshift range $0.4 \leq z\leq 0.6$.

\begin{figure}
\includegraphics[width = 3.2in]{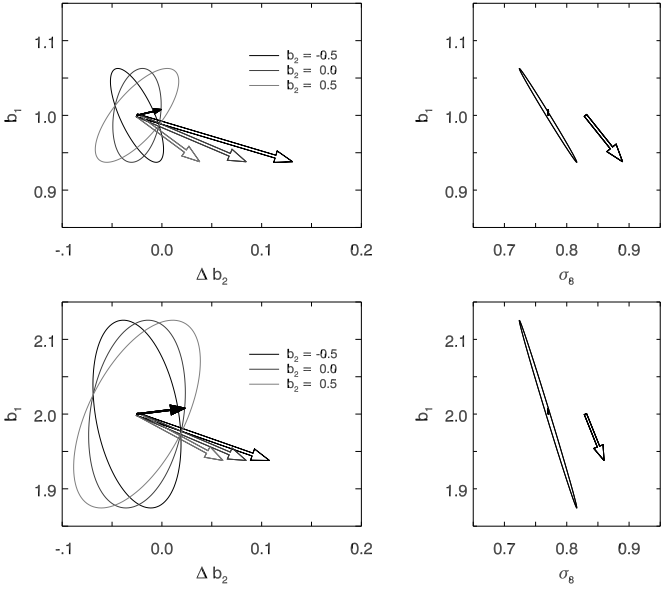}
\caption{Systematic errors induced by intrinsic alignment. Ellipses show $95\%$ C.L. statistical errors on parameter estimates in a DES-like surveys for a fiducial model with $\sigma_8  =0.8$, for a galaxy population with $b_1 = 1$ (top panels) or $b_1 = 2$ (bottom panel) and $b_2\in \{-0.5,0,0.5\}$. Open/ filled arrows illustrate the systematic parameter shift induced by a quadratic/ linear intrinsic alignment contamination.}
\label{fig:bias}
\end{figure}

The systematic error on the bispectrum, $\vec{\Delta \mathcal B}$, due to linear or quadratic alignment is modeled by the line of sight projection 
(Eq.~\ref{eq:Limber}) of the tidal alignment contaminations (Eqs.~\ref{eq:D_BLA}, \ref{eq:B_QA}) calculated in Sect.~\ref{sec:TA}. We set 
$\vec{\Delta 
\mathcal P} = 0$ for the quadratic alignment model as the first correction to the power spectrum is third order in the density contrast. In agreement 
with our findings from Eq.~(\ref{eq:bias_LA}), the systematic error induced by linear alignment on the galaxy power spectrum is given by 
\citep[cf.][]{Hirata09}
\be
\Delta P_{\mr g}^{\mr{LA}}(k_{\perp}) = \left[\left(b_1 - \frac{A_1}3\right)^2 - b_1^2\right] P_{\mr g}(k_{\perp}),
\ee
where we have restricted $\bmath k$ to be orthogonal to the line of sight as only these modes survive the Limber approximation.

Figure~\ref{fig:bias} shows the marginalized Fisher matrix estimates of statistical parameter errors ($95\%$ C.L.) obtained with our fictitious survey in the absence of an intrinsic alignment, and the systematic bias induced by a linear  or quadratic alignment contamination.

The systematic bias induced by a linear alignment contamination (solid arrows) we find through the Fisher matrix analysis (Eq.~\ref{eq:Fisher_bias}) is 
in 
agreement with the analytic result (Eq.~\ref{eq:bias_LA}). The parameter bias on $b_1$ is independent of the value of $b_2$ assumed in the fiducial model and 
the 
solid arrows of different color are indistinguishable. Assuming a normalization of $A_1 = -0.024$ as discussed in Sect.~\ref{sec:TA}, the systematic 
error on $b_2$ is comparable to the $95\%$ C.L. statistical error for $b_2$ in our survey. The systematic error on $b_1$ caused by  the linear 
alignment model is smaller, but may still
be important if many photo-$z$ slices are used 
in the parameter analysis.  In the limit of our toy model, the effect of linear alignment on the angular galaxy power spectrum and bispectrum 
is fully described by a systematic error in the linear and non-linear bias parameter (Eq.~\ref{eq:bias_LA}) and it has no effect on measurements of 
$\sigma_8$.

The strength of the quadratic alignment contamination depends on triangle shape and size; it is {\em not} well described by a rescaling of the galaxy 
bias 
parameters. Hence the Fisher matrix estimates for the systematic parameter errors depend on the binning scheme and range of scales adopted in the 
analysis. For our choice 
of 20 equidistant bins per angular frequency, and with the range of scales of 0.04--0.2$h\,$Mpc$^{-1}$, we a systematic shift towards larger non-linear 
bias $b_2$ 
and smaller $b_1$. The latter is degenerate between $b_1$ and $\sigma_8$. The plot illustrates a quadratic alignment contamination with normalization 
$A_2 = 1$. As can been seen from Eqs.~(\ref{eq:B_QA}, \ref{eq:Fisher_bias}), the systematic bias is linear in $A_2$, and it reverses sign if $A_2<0$. 
While exact form of the systematic 
error caused by the toy model for quadratic alignment depends on a number of parameters, it may cause a significant contamination in our fictitious 
survey if $|A_2| \gsim 0.5$, or if (as we expect) multiple photo-$z$ slices are used to reduce statistical errors.

\section{Discussion}
\label{sec:theend}

Using simple toy models for intrinsic alignment and the local bias approximation we have analyzed the effect of tidal alignment on the galaxy 
bispectrum. If the orientation of galaxies depends on the surrounding tidal field, and if the detection probability for galaxies is orientation 
dependent, the observed clustering of galaxies is modified by tidal alignments. This astrophysical contaminant can introduce systematic errors to 
parameters derived from the bispectrum.

\begin{figure}
\includegraphics[width = 3.2in]{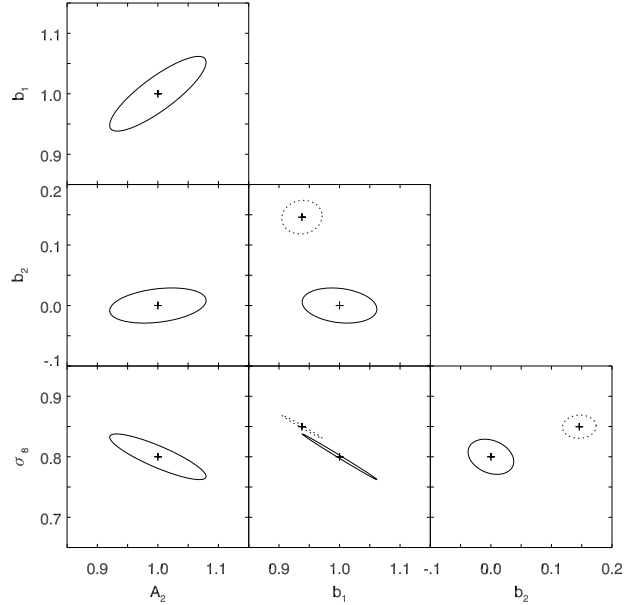}
\caption{Removal of quadratic alignment bias. Dotted ellipses show the biased parameter estimates and their $95\%$ contour regions in the presence of quadratic alignment contamination with $A_2 = 1$ which is unaccounted for in the analysis. The solid ellipses illustrate the $95\%$ contour regions of the unbiased parameter estimates in an analysis which includes a quadratic alignment contamination and marginalizes over $A_2$.}
\label{fig:A2}
\end{figure}

A toy model for linear alignments \citep{CKB}, which is based on the assumption that tidal fields elongate/compress haloes and thus determine galaxy 
shapes, results in a rescaling of linear and non-linear galaxy bias parameters that is proportional to the strength of the halo shape distortion. The 
presence of this systematic error in the observed galaxy bias measurements cannot be detected from projected clustering data as the strength of the 
alignment contamination is completely degenerate with the unobservable true bias parameters and outside information will be necessary to remove it. 
Normalizing the strength of the linear tidal alignment toy model to measurements of intrinsic alignments in weak lensing observations, we find that 
linear alignment may introduce systematic errors to galaxy bias measurements at the percent level (again using only the real-space observables), and 
thus will likely not be significant.

Using a simple model for quadratic alignment based on galaxy spin correlations in linear tidal torque theory we calculate a systematic contamination 
which modifies the shape of the galaxy bispectrum. Depending on survey characteristics, we find that quadratic alignment may introduce significant 
systematic errors to the galaxy bias parameters and the normalization of the power spectrum derived from the angular galaxy bispectrum. As the 
quadratic alignment contamination has different shape than the galaxy bispectrum, one can include a model for the contamination in the analysis and 
marginalize over its normalization. Figure~\ref{fig:A2} illustrates how such a marginalization may remove the systematic bias at the cost of larger 
statistical errors. The biased data points and contour levels (dashed lines) are taken from Fig.~\ref{fig:bias} for a fiducial model with $b_1 = 1$ and 
$b_2 = 0$. The new statistical errors including marginalization over $A_2$ are calculated by adding $A_2$ as a nuisance parameter and including the 
contamination signal in the fiducial model of the Fisher matrix analysis ($\vec{\mathcal B} \rightarrow \vec{\mathcal B}+\vec{\Delta \mathcal B}$ in Eq.~\ref{eq:fisher}).

This analysis lives in the weakly nonlinear regime to enable the use of simple models for linear and quadratic alignment. As the information content of 
the bispectrum increases dramatically with the maximal spatial frequency that is included in an analysis, any realistic analysis will have extend well 
into the quasilinear regime. While models from the redshift space bispectrum on these scales \citep{S+08} approach the required accuracy for such 
analyses, the treatment of tidal alignments including the non-Gaussian nature of the angular moment distribution and non-linear stages of galaxy 
formation requires further work.

\section*{Acknowledgements}

E.K. and C.M.H. are supported by the U.S. Department of Energy (DE-FG03-92-ER40701) and the National Science Foundation (AST-0807337).
C.M.H. is supported by the Alfred P. Sloan Foundation.

\appendix

\section{Tree-level galaxy trispectrum}

To calculate the tree-level matter trispectrum we need to consider the density contrast to third order as the tree-level Trispectrum splits into two 
types of connected terms, $\ensav{\tilde \delta^{(1)}\tilde \delta^{(1)}\tilde \delta^{(2)}\tilde \delta^{(2)}}_{\mr c}$ and $\ensav{\tilde \delta^{(1)}\tilde \delta^{(1)}\tilde \delta^{(1)}\tilde \delta^{(3)}}_{\mr c}$. 
The third order density contrast is given by \citep{Fry84}
\beq
\nonumber \tilde \delta^{(3)}(\bmath k) \!\! &=& \!\! \int\frac{\mr d^3 \bmath k_1}{(2\pi)^3} \int\frac{\mr d^3 \bmath k_2}{(2\pi)^3}F_3\left(\bmath 
k_1,\bmath k_2, \bmath k - \bmath k_1 - \bmath k_2\right)\\
&&\!\!
\times \tilde\delta^{(1)}(\bmath k_1)\tilde\delta^{(1)}(\bmath k_2)\tilde\delta^{(1)}( \bmath k - \bmath k_1 - \bmath k_2)\;,
\eeq
with the third order coupling function $F_3$.
One finds for the matter trispectrum 
\beq
\nonumber (2\pi)^3 \delta_{\rm D}(\bmath k_{1234}) T_{\mr{pt}}(\bmath k_1,\bmath k_2,\bmath k_3,\bmath k_4) \approx ~~~~~~~~~~~~~~ && \\
\nonumber \ensav{\tilde \delta^{(1)}(\bmath k_1)\tilde \delta^{(1)}(\bmath k_2)\tilde \delta^{(1)}(\bmath k_3)\tilde \delta^{(3)}(\bmath k_4)}
\!\!\!\! &+& \!\!\!\! 3\;\mr{perm.}\\
 +\ensav{\tilde \delta^{(1)}(\bmath k_1)\tilde \delta^{(1)}(\bmath k_2)\tilde \delta^{(2)}(\bmath k_3)\tilde \delta^{(2)}(\bmath k_4)}
\!\!\!\! &+& \!\!\!\! 5\;\mr{perm.}
\eeq
After some algebra one obtains
\beq
\nonumber  T_{\mr{pt}}(\bmath k_1,\bmath k_2,\bmath k_3,\bmath k_4)
\!\! &=& \!\! 6F^{\mr s}_3(\bmath k_1,\bmath k_2,\bmath k_3) P(k_1)P(k_2)P(k_3)+ 3\;\mr{perm.}\\
\nonumber && \!\! + 4[P(k_{13})F_2(\bmath k_1, -\bmath k_{13})F_2(\bmath k_2, \bmath k_{13})\\
\nonumber && \!\!
+P(k_{23})F_2(\bmath k_1, \bmath k_{23})F_2(\bmath k_2, -\bmath k_{23})]\\
&& \!\! \times P(k_1)P(k_2) +5\;\mr{perm.}
\eeq
If one assume the third order galaxy bias ($b_3$) to be zero, two types of additional terms containing the quadratic galaxy bias contribute to the 
galaxy trispectrum, $\ensav{b_1\tilde \delta^{(1)}\;b_1\tilde \delta^{(1)}\;b_1 \tilde \delta^{(2)}\; b_2\tilde \delta^{(1)}\otimes\tilde 
\delta^{(1)}}_{\mr c}$ and $\ensav{b_1\tilde \delta^{(1)}\;b_1\tilde \delta^{(1)}\;b_2\tilde \delta^{(1)}\otimes\tilde \delta^{(1)}\; b_2\tilde 
\delta^{(1)}\otimes\tilde \delta^{(1)}}_{\mr c}$. Hence our model for the galaxy trispectrum is given by
\beq
\nonumber T_{\mr{gal}}(\bmath k_1,\bmath k_2,\bmath k_3,\bmath k_4)
\!\! &\approx& \!\! b_1^4 T_{\mr{pt}}(\bmath k_1,\bmath k_2,\bmath k_3,\bmath k_4) \\
\nonumber && \!\! +2 b_1^3 b_2 P(k_1)P(k_2)[P(k_{13})F_2(\bmath k_1, -\bmath k_{13}) \\
\nonumber && \!\!
+P(k_{24})F_2(\bmath k_2, -\bmath k_{23})] + 5\;\mr{perm.}\\
\nonumber && \!\!
+4 b_1^2b_2^2 P(k_1)P(k_2)\left[P(k_{13})+P(k_{23})\right] \\
&& \!\!
+ 5\;\mr{perm.}
\label{eq:T_g}
\eeq

\label{lastpage}

\begin{thebibliography}{}

\bibitem[\protect\citeauthoryear{{Amara} \& {R{\'e}fr{\'e}gier}}{2008}]{AR08}
{Amara} A., {R{\'e}fr{\'e}gier} A., 2008, \mnras, 391, 228

\bibitem[\protect\citeauthoryear{{Berlind} \& {Weinberg}}{2002}]{BW02}
{Berlind} A.~A., {Weinberg} D.~H., 2002, \apj, 575, 587

\bibitem[\protect\citeauthoryear{{Bernardeau} {et~al.}}{2002}]{Bernardeau02}
{Bernardeau} F., {Colombi} S., {Gazta{\~n}aga} E., {Scoccimarro}, R.,
  2002, \physrep, 367, 1

\bibitem[\protect\citeauthoryear{{Binggeli}}{1982}]{B82}
{Binggeli} B., 1982, \aap, 107, 338

\bibitem[\protect\citeauthoryear{{Buchalter} {et~al.}}{2000}]{BKJ00}
{Buchalter} A., {Kamionkowski} M., {Jaffe} A.~H., 2000, \apj, 530, 36

\bibitem[\protect\citeauthoryear{{Catelan} {et~al.}}{2001}]{CKB}
{Catelan} P., {Kamionkowski} M., {Blandford} R.~D., 2001, \mnras, 320, L7

\bibitem[\protect\citeauthoryear{{Cooray}}{2004}]{C04}
{Cooray} A., 2004, \mnras, 348, 250

\bibitem[\protect\citeauthoryear{{Cooray} \& {Sheth}}{2002}]{CS02}
{Cooray} A., {Sheth} R., 2002, \physrep, 372, 1

\bibitem[\protect\citeauthoryear{{Crittenden} {et~al.}}{2001}]{CNPT01}
{Crittenden} R.~G., {Natarajan} P., {Pen} U.-L., {Theuns} T., 2001, \apj,
  559, 552

\bibitem[\protect\citeauthoryear{{Dolney} {et~al.}}{2006}]{DTJ06}
{Dolney} D., {Jain} B., {Takada} M., 2006, \mnras, 366, 884

\bibitem[\protect\citeauthoryear{{Doroshkevich}}{1970}]{D70}
{Doroshkevich} A.~G., 1970, Astrophysics, 6, 320

\bibitem[\protect\citeauthoryear{{Faltenbacher} {et~al.}}{2007}]{FLM07}
{Faltenbacher} A., {Li} C., {Mao} S. {et~al.}, 2007, \apjl, 662, L71

\bibitem[\protect\citeauthoryear{{Faltenbacher} {et~al.}}{2009}]{FLW09}
{Faltenbacher} A., {Li} C., {White} S.~D.~M. {et~al.}, 2009, Research in
  Astronomy and Astrophysics, 9, 41

\bibitem[\protect\citeauthoryear{{Feldman} {et~al.}}{2001}]{FFS01}
{Feldman} H.~A., {Frieman} J.~A., {Fry} J.~N., {Scoccimarro} R., 2001,
  \prl, 86, 1434

\bibitem[\protect\citeauthoryear{{Fry}}{1984}]{Fry84}
{Fry} J.~N., 1984, \apj, 279, 499

\bibitem[\protect\citeauthoryear{{Fry}}{1994}]{F94}
{Fry} J.~N., 1994, \prl, 73, 215

\bibitem[\protect\citeauthoryear{{Fry} \& {Gazta{\~n}aga}}{1993}]{FG93}
{Fry} J.~N., {Gazta{\~n}aga} E., 1993, \apj, 413, 447

\bibitem[\protect\citeauthoryear{{Fry} \& {Thomas}}{1999}]{FT99}
{Fry} J.~N., {Thomas} D., 1993, \apj, 524, 591

\bibitem[\protect\citeauthoryear{{Guo} \& {Jing}}{2009}]{GJ09B}
{Guo} H., {Jing} Y.~P., 2009, \apj, 698, 479

\bibitem[\protect\citeauthoryear{{Guo} \& {Jing}}{2009}]{GJ09}
{Guo} H., {Jing} Y.~P., 2009, \apj, 702, 425

\bibitem[\protect\citeauthoryear{{Hatton} \& {Cole}}{1998}]{HC98}
{Hatton} S., {Cole} S., 1998, \mnras, 296, 10

\bibitem[\protect\citeauthoryear{{Heavens} {et~al.}}{1998}]{HMV98}
{Heavens} A.~F., {Matarrese} S., {Verde} L., 1998, \mnras, 301, 797

\bibitem[\protect\citeauthoryear{{Hirata}}{2009}]{Hirata09}
{Hirata} C.~M., 2009, \mnras, 399, 1074

\bibitem[\protect\citeauthoryear{{Hirata} {et~al.}}{2007}]{H07}
{Hirata} C.~M., {Mandelbaum} R., {Ishak} M. {et~al.}, 2007, \mnras, 381,
  1197

\bibitem[\protect\citeauthoryear{{Hoyle}}{1949}]{Hoyle}
{Hoyle} F., 1949, \mnras, 109, 365

\bibitem[\protect\citeauthoryear{{Hui} \& {Zhang}}{2008}]{HZ08}
{Hui} L., {Zhang} J., 2008, \apj, 688, 742

\bibitem[\protect\citeauthoryear{{Huterer} {et~al.}}{2006}]{H06}
{Huterer} D., {Takada} M., {Bernstein} G., {Jain} B., 2006, \mnras, 366,
  101

\bibitem[\protect\citeauthoryear{{Jackson}}{1972}]{Jackson72}
{Jackson} J., 1972, \mnras, 156, 1

\bibitem[\protect\citeauthoryear{{Jing} \& {B{\"o}rner}}{2004}]{JB04}
{Jing} Y.~P., {B{\"o}rner} G., 2004, \apj, 607, 140

\bibitem[\protect\citeauthoryear{{Joachimi} {et~al.}}{2009}]{JXS10}
{Joachimi} B., {Shi} X., {Schneider} P., 2009, \aap, 508, 1193

\bibitem[\protect\citeauthoryear{{Kaiser}}{1984}]{K84}
{Kaiser} N., 1984, \apjl, 284, L9

\bibitem[\protect\citeauthoryear{{Kaiser}}{1987}]{K87}
{Kaiser} N., 1987, \mnras, 227, 1

\bibitem[\protect\citeauthoryear{{Kaiser}}{1992}]{Kaiser92}
{Kaiser} N., 1992, \apj, 388, 272

\bibitem[\protect\citeauthoryear{{Kayo} {et~al.}}{2004}]{KSN04}
{Kayo} I., {Suto} Y., {Nichol} R.~C. {et~al.}, 2004, \pasj, 56, 415

\bibitem[\protect\citeauthoryear{{Komatsu} {et~al.}}{2010}]{WMAP7}
{Komatsu} E., {Smith} K.~M., {Dunkley} J. {et~al.}, 2010,
  \apjs, submitted, preprint: arXiv:1001.4538

\bibitem[\protect\citeauthoryear{{Kulkarni} {et~al.}}{2007}]{KNS07}
{Kulkarni} G.~V., {Nichol} R.~C., {Sheth} R.~K. {et~al.}, 2007, \mnras, 378,
  1196

\bibitem[\protect\citeauthoryear{{Lee} \& {Pen}}{2000}]{LP00}
{Lee} J., {Pen} U.-L., 2000, \apj, 532, L5

\bibitem[\protect\citeauthoryear{{Lee} \& {Pen}}{2007}]{LP07}
{Lee} J., {Pen} U.-L., 2007, \apj, 670, 1

\bibitem[\protect\citeauthoryear{{Mandelbaum} {et~al.}}{2009}]{MWiggleZ}
{Mandelbaum} R., {Blake} C., {Bridle} S. {et~al.}, 2009, \mnras, submitted,
  preprint: arXiv:0911.5347

\bibitem[\protect\citeauthoryear{{Manera} \& {Gazta{\~n}aga}}{2009}]{MG09}
{Manera} M., {Gazta{\~n}aga} E. 2009, \mnras, submitted,
  preprint: arXiv:0912.0446

\bibitem[\protect\citeauthoryear{{Mar{\'{\i}}n} {et~al.}}{2008}]{Marin08}
{Mar{\'{\i}}n} F.~A., {Wechsler} R.~H., {Frieman} J.~A., {Nichol} R.~C.
  2008, \apj, 672, 849

\bibitem[\protect\citeauthoryear{{McDonald}}{2006}]{McDonald}
{McDonald} P., 2006, \prd, 74, 103512

\bibitem[\protect\citeauthoryear{{McDonald} \& {Roy}}{2009}]{McDonald2}
{McDonald} P., {Roy} A., 2009, \jcap, 08, 020

\bibitem[\protect\citeauthoryear{{Nichol} {et~al.}}{2006}]{NSS06}
{Nichol} R.~C., {Sheth} R.~K., {Suto} Y. {et~al.}, 2006, \mnras, 368, 1507

\bibitem[\protect\citeauthoryear{{Niederste-Ostholt} {et~al.}}{2010}]{NSD10}
{Niederste-Ostholt} M., {Strauss} M.~A., {Dong} F. {et~al}, 2010, \mnras, accepted, preprint: arXiv:1003.0322

\bibitem[\protect\citeauthoryear{{Nock} {et~al.}}{2010}]{NPR10}
{Nock} K., {Percival} W.~J., {Ross} A.~J., 2010, \mnras, submitted,
  preprint: arXiv:1003.0896

\bibitem[\protect\citeauthoryear{{Padmanabhan} {et~al.}}{2007}]{PSS07}
{Padmanabhan} N., {Schlegel} D.~J., {Seljak} U. {et~al.}, 2007, \mnras, 378,
  852

\bibitem[\protect\citeauthoryear{{Peacock} {et~al.}}{2001}]{P01}
{Peacock} J., {Cole} S., {Norberg} P. {et~al.}, 2001, Nature, 410, 169

\bibitem[\protect\citeauthoryear{{Peebles}}{1969}]{P69}
{Peebles} P.~J.~E., 1969, \apj, 155, 393

\bibitem[\protect\citeauthoryear{{Reid} {et~al.}}{2009}]{Reid09}
{Reid} B.~A., {Spergel} D.~N., {Bode} P., 2009, \apj, 702, 249

\bibitem[\protect\citeauthoryear{{Sch{\"a}fer}}{2009}]{BMS09}
{Sch{\"a}fer} B.~M., 2009, Int. J. Mod. Phys. D, 18, 173

\bibitem[\protect\citeauthoryear{{Sciama}}{1955}]{Sciama}
{Sciama} D., 1955, \mnras, 115, 3

\bibitem[\protect\citeauthoryear{{Scoccimarro} {et~al.}}{1999}]{SCF99}
{Scoccimarro} R., {Couchman} H.~M.~P., {Frieman} J.~A., 1999, \apj, 517,
  531

\bibitem[\protect\citeauthoryear{{Scoccimarro} {et~al.}}{2001}]{SFF01}
{Scoccimarro} R., {Feldman} H.~A., {Fry} J.~N., {Frieman} J.~A., 2001,
  \apj, 546, 652

\bibitem[\protect\citeauthoryear{{Sefusatti} {et~al.}}{2006}]{SCPS06}
{Sefusatti} E., {Crocce} M., {Pueblas} S. {et~al} , 2006, \prd,
  74, 023522

\bibitem[\protect\citeauthoryear{{Seljak}}{2000}]{Seljak00}
{Seljak} U., 2000, \mnras, 318, 203

\bibitem[\protect\citeauthoryear{{Seljak} \& {Zaldarriaga}}{1996}]{CMBFAST}
{Seljak} U., {Zaldarriaga} M., 1996, \apj, 469, 437

\bibitem[\protect\citeauthoryear{{Sheth} \& {Tormen}}{1999}]{ShethTormen}
{Sheth} R., {Tormen} G., 1999, \mnras, 308, 119

\bibitem[\protect\citeauthoryear{{Smith} {et~al.}}{2008}]{S+08}
{Smith} R.~E., {Sheth} R.~K., {Scoccimarro} R., 2008, \prd, 78, 023523

\bibitem[\protect\citeauthoryear{{Tegmark} {et~al.}}{2004}]{Tegmark04}
{Tegmark} M., {Blanton} M.~R., {Strauss} M.~A. {et~al.}, 2004, \apj, 606,
  702

\bibitem[\protect\citeauthoryear{{van den Bosch} {et~al.}}{2002}]{vandenBosch}
{van den Bosch} F., {Abel} T., {Croft} R., {Hernquist} L., {White} S.~D.~M., 2002,
  \apj, 576, 21

\bibitem[\protect\citeauthoryear{{Verde} {et~al.}}{1998}]{VHM98}
{Verde} L., {Heavens} A., {Matarrese} S., {Moscardini} L., 1998,
  \mnras, 300, 747

\bibitem[\protect\citeauthoryear{{Verde} {et~al.}}{2000}]{V00}
{Verde} L., {Heavens} A., {Matarrese} S., 2000, \mnras, 318, 584

\bibitem[\protect\citeauthoryear{{Verde} {et~al.}}{2002}]{VHP02}
{Verde} L., {Heavens} A., {Percival} W. {et~al.}, 2002, \mnras, 335,
  432

\bibitem[\protect\citeauthoryear{{White}}{1984}]{W84}
{White} S.~D.~M., 1984, \apj, 286, 38

\end{thebibliography}
\end{document}